\documentclass[aps,superscriptaddress,twocolumn,prb]{revtex4-1}

\usepackage{chemformula} % Formula subscripts using \ch{}
\usepackage[T1]{fontenc} % Use modern font encodings
\usepackage{amsfonts,latexsym}
\usepackage{amsmath}
\usepackage{graphicx}
\usepackage{siunitx}
\newcommand{\la}{\langle}
\newcommand{\ra}{\rangle}
\newcommand{\singbu}{$^1\!$B$_\textrm{u}${ }}
\newcommand{\singag}{$^1\!$A$_\textrm{g}${ }}
\newcommand{\tripbu}{$^3\!$B$_\textrm{u}${ }}

\begin{document}

\author{Igor Lyskov}
\email{igor.lyskov@rmit.edu.au}
%\phone{+61 3 9925 1647}
%\fax{+61 3 9926 5290}
\affiliation{ARC Centre of Excellence in Exciton Science}
\affiliation{School of Science, RMIT University, Melbourne, 3000, Australia}
\author{Egor Trushin}
\affiliation{Lehrstuhl f\"ur Theoretische Chemie, Universit\"at Erlangen-N\"urnberg, Egerlandstr. 3, D-91058 Erlangen, Germany}
\author{Ben Q. Baragiola}
\affiliation{School of Science, RMIT University, Melbourne, 3000, Australia}
\affiliation{ARC Centre of Excellence for Quantum Computation and Communication Technology}
\author{Timothy W. Schmidt}
\affiliation{ARC Centre of Excellence in Exciton Science}
\affiliation{School of Chemistry, UNSW Sydney, NSW 2052, Australia}
\author{Jared H. Cole}
\affiliation{ARC Centre of Excellence in Exciton Science}
\affiliation{School of Science, RMIT University, Melbourne, 3000, Australia}
\author{Salvy P. Russo}
\affiliation{ARC Centre of Excellence in Exciton Science}
\affiliation{School of Science, RMIT University, Melbourne, 3000, Australia}

\title{First-Principles Calculation of Triplet Exciton Diffusion in Crystalline Poly({\it p}-phenylene vinylene)}

%\begin{tocentry}
%\centering
 % \includegraphics{toc}
%\end{tocentry}

%% if an abstract is not used by the targetal.
%%%%%%%%%%%%%%%%%%%%%%%%%%%%%%%%%%%%%%%%%%%%%%%%%%%%%%%%%%%%%%%%%%%%%

\begin{abstract}
\begin{minipage}[t]{0.3\linewidth}
In this article we present multiscale modelling of triplet exciton energy migrating through the archetypical poly({\it p}-phenylene vinylene) polymer (PPV) in the crystal phase. We combine electronic structure calculations with coupled exciton-nuclear quantum dynamics in order to parameterize exciton evolution in J- and H-aggregate configurations. We then apply this parameterization to a master-equation approach to describe transport at the nanoscale. 
\end{minipage}
\vtop{%
  \vskip-1ex
  \hbox{%
    \includegraphics[width=0.50\linewidth]{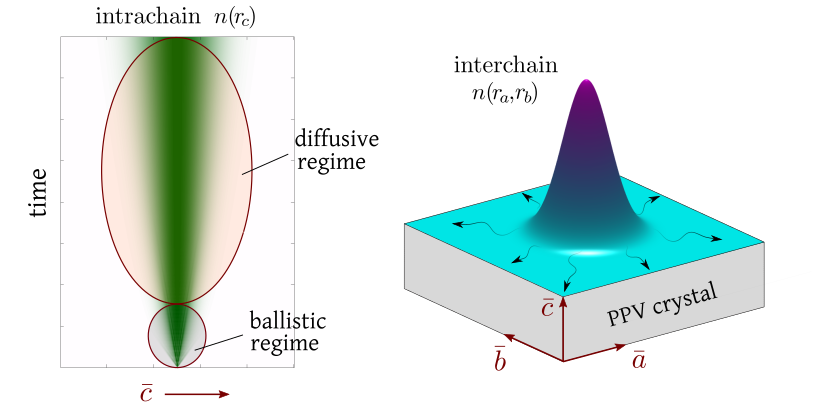}%
  }%
}%

\vskip1ex
\begin{minipage}[t]{0.8\linewidth}
We find that triplet transport is characterized by two remarkably different components: fast and coherent intrachain, slow and incoherent interchain. Energy migration along the polymer backbone is accompanied by coherent superpositions developing  between neighboring sites in the first 20 fs, however no interchain coherence develops. The nonequilibrium exciton density exhibits an initial ultrafast ballistic spread followed by normal diffusive propagation.  At room temperature the diffusion coefficients along respective interchain axes are found to be $D_a=2.48\cdot10^{-2}$ cm$^2$s$^{-1}$ and $D_b=4.18\cdot10^{-2}$ cm$^2$s$^{-1}$, and $D_c=3.03$ cm$^2$s$^{-1}$ along the fast axis.
\end{minipage}
\end{abstract}

\maketitle


\section{Introduction}

A fundamental understanding of nonequilibrium excitonic systems can pave the way towards rational design of novel optoelectronic technologies. Systematic characterization of the physical processes, at each stage from the initial exciton generation until recombination, is a cornerstone for rational fabrication strategies. In particular, understanding and controlling nanoscale transport is one of the crucial facets in functional materials design~\cite{Menke2014,Ostroverkhova2016,Tamai2015,Brixner2017,Yost2012}.

For the case of organic materials, triplet excitons offer the advantage of typically long temporal lifetimes for energy harvesting. Despite their low mobility compared to singlets, triplets can have large diffusion lengths, spanning over sub-mm lengthscale, as reported in polyacene crystals~\cite{Irkhin2011,Williams1966,Najafov2010,Akselrod2014,Wan2015}. However, the dark nature of a triplet alongside the various exciton transformations, e.g. nonradiative recombination, spin interconversion and dissociation into charge carriers, hamper experimental analysis of triplet transport. Moreover, possible crystal imperfections along with low rigidity of organic materials inevitably influence the accuracy and refinement of time-dependent measurements~\cite{Grieco2016}. This motivates a theoretical characterization of space-time evolution of the triplet exciton in molecular crystals.

A computational approach to a quantum particle migrating through the bulk molecular material spans various strategies, from semiclassical Marcus theory~\cite{Yost2012,Stehr2014,Kranz2016} and surface hopping~\cite{Kranz2016,Giannini2018,Wang2013} to coupled electron-nuclear dynamics~\cite{Arago2016,Fornari2016,Troisi2006,Troisi2007}. It has been shown that transport properties of excitons in molecular solids display anisotropy with respect to the principal crystal axes~\cite{Arago2016,Irkhin2011,Akselrod2014,Stehr2014}. Conjugated polymeric solids are intrinsically nonhomogeneous materials~\cite{Spano2014,Tamai2015}. This necessitates a thorough assessment of the two essential components of transient dynamics, i.e. intrachain and interchain. Energy migration across adjacent chains is conventionally described in the framework of F\"orster and Dexter theory~\cite{Ostroverkhova2016,Brixner2017}. On the other hand, due to the effects of $\pi$-conjugation, the energy transfer along the backbone chain cannot be described in the limits of weak-coupling theory in general.  It is argued that exciton evolution in polymers proceeds in two different regimes: (i) incoherent hopping for the case of H-aggregate (interchain), and (ii) coherent wave-like propagation through the supramolecular J-aggregate (intrachain).

\begin{figure}[tb]
\centering
 \includegraphics[width=0.6\linewidth]{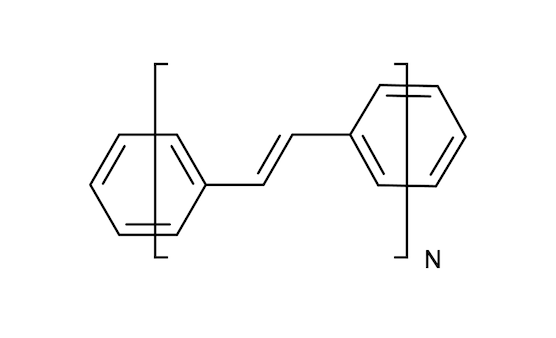}
 \caption{Chemical structure of PPV$_\textrm{N}$ oligomers.}
 \label{fig:ppvn}
\end{figure}

Many recent theoretical studies have been devoted to the dynamical properties of singlet exciton in polymers~\cite{Tozer2012,Barford2012,Tozer2015,Mannouch2018}, including sophisticated multidimensional dynamical treatments~\cite{Binder2013,Binder2018}. Although triplets serve as productive energy carriers in polymer matrices utilized in modern energy upconversion devices~\cite{Simon2012,Jankus2013,Raisys2018}, less is known about their transient transport nature. In this article we use a multiscale approach to characterize the intra- and interchain mobility of triplet excitons in an archetypical polymer PPV (Figure~\ref{fig:ppvn}) in the condensed phase. We employ a model vibronic Hamiltonian, parameterized as a function of on-site and junction nuclear vibrations, for full quantum dynamical treatment of the triplet exciton migration between neighboring PPV fragments. The vibronic parameterization is entirely based on high-level electronic structure calculations. From the transient exciton evolution, we deduce the system-bath parameters which are used in the Lindblad master equation describing energy transfer at the nanoscale.

\section{Electronic properties of excitons in PPV}

In the absence of external fields,  time evolution of excitons in a material is driven by electronic structure properties of this material. The solution of the time-independent Schr\"odinger equation provides a set of stationary states, which can be effectively mapped onto a subsystem-level by dividing the polymer into small structural domains. Considering a polymer chain as a set of equivalent fragments (sites) allows us to create a nonequilibrium initial condition for a time-dependent treatment and to elucidate exciton migration in PPV. In this section we discuss the excited-state properties of PPV in the crystalline phase and in short oligomers. Based on the computed excited-state spectrum we extract the on-site energy of monomers and interfragment coupling for J- and H-aggregates. Further, we describe important nuclear vibrations which are appreciably coupled to the on-site Frenkel excitons.

\subsection{Solid state properties}

Density functional theory (DFT)~\cite{Dreizler1990}, the Green's function method (GW)~\cite{Onida2002} and Bethe-Salpeter equation (BSE)~\cite{Onida2002} calculations with periodic boundary conditions were performed using the \textsc{vasp}~\cite{Kresse1993,Kresse1994,Kresse1996,Kresse1996a} package with the projector augmented wave method~\cite{Bloechl1994,Kresse1999}. The exchange-correlation functional due to Perdew-Burke-Ernzerhof (PBE)~\cite{Perdew1996} was employed in DFT calculation to relax the structural parameters and to provide a starting reference electronic structure for the GW calculations. We optimized the geometry using a plane-wave energy cutoff of 400 eV and a regular $\Gamma$-point-centered grid of 8$\times$8$\times$8 \textbf{k}-points to sample the Brillouin zone.

\begin{figure}[tb]
\centering
 \includegraphics[width=\linewidth]{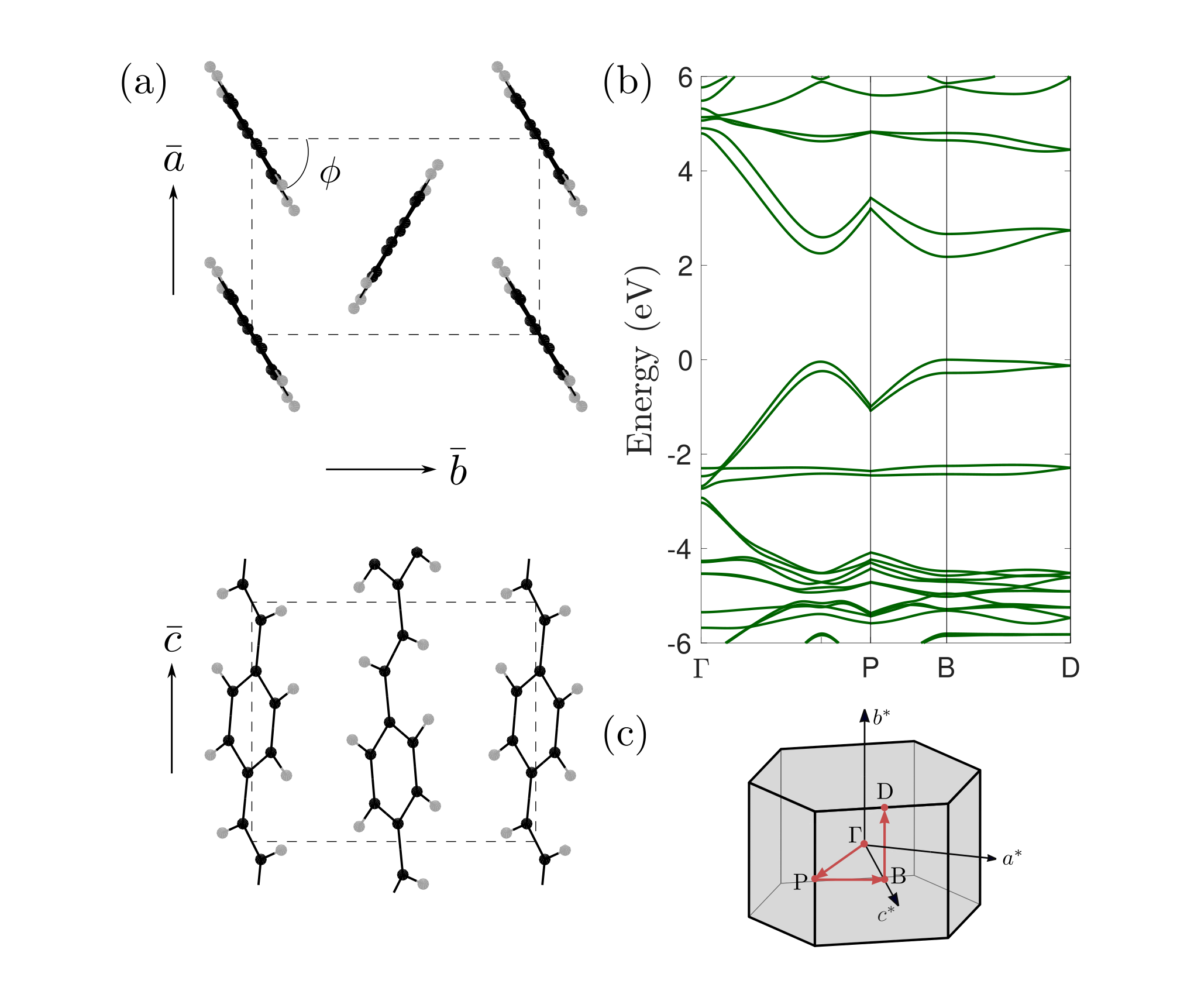}
 \caption{(a) Chain arrangement in the PPV crystal with monoclinic unit cell. (b) The GW$_0@$PBE electronic band structure with band gap of 2.18 eV. Energy offset is the top HOMO position. (c) The Brillouin zone of the crystal with high-symmetry points. The chain direction is along $\Gamma\textrm{P}$ line.}
 \label{fig:bands}
\end{figure}

The PPV is crystallized in a monoclinic structure with herringbone packing of the polymer chains. The unit cell of the crystal with P$2_1/n$ space symmetry group contains two symmetry equivalent units with setting angle $\phi=57^\circ$ as shown in Figure~\ref{fig:bands}. The resulting lattice parameters are $a$=6.12 {\AA}, $b$=7.94 {\AA}, $c$=6.69 {\AA}, and $\hat{\bf{ac}}=119^\circ$, which agree very well with the X-ray diffraction data from Ref.~\citenum{Granier1986} ($a$=6.05 {\AA}, $b$=7.90 {\AA},  $c$=6.58 {\AA}, and $\hat{\bf{ac}}=123^\circ$) and Ref.~\citenum{Chen1990} ($a$=6.05 {\AA}, $b$=8.07 {\AA}, $c$=6.54 {\AA}, and $\hat{\bf{ac}}=123^\circ$).

Based on the optimized crystal structure, we evaluate the band structure and excitonic energy levels using a 6$\times$6$\times$6 \textbf{k}-point grid and a plane-wave energy cutoff of 500 eV. The GW$_0$ band structure along the $\Gamma$-P-B-D path of the Brillouin zone,  obtained using the \textsc{wannier90} code~\cite{Mostofi2014}, is shown in Figure~\ref{fig:bands}. The PPV crystal possesses a direct band gap, located at the B-point of the Brillouin zone in both DFT-PBE and GW calculations (see supplemental information for the DFT bands). The DFT-PBE provides the band gap of 0.87 eV, which substantially deviates from the GW$_0$ value of 2.18 eV. Furthermore, the GW$_0$ method increases the electronic bandwidth and the magnitude of band splittings. The largest dispersion occurs along the $\Gamma$-P path, which coincides with the direction of the individual chains, implying the dominant role of intrachain transport in solid state PPV.

In contrast to conventional semiconductors, organic crystals possess large exciton binding energy that gives a substantial difference between optical and electronic band gaps. We determine the excitonic energies by solving the BSE within the Tamm-Dancoff approximation~\cite{Onida2002}. The GW$_0$-BSE method places the S$_1$ and S$_2$ energies at 1.84 and 1.95 eV, respectively. Due to the photophysical properties of H-aggregates, the lowest singlet state is optically dark, whilst the S$_2$ is the first optically active direct excitation. The exchange interaction between the $\pi$-electron and $\pi$-hole substantially stabilizes triplet excitons compared to singlets. The GW$_0$-BSE energies for T$_1$ and T$_2$ are 1.52 and 1.57 eV, which is in line with experimental values 1.45-1.55 eV~\cite{Colaneri1990,Oesterbacka2003}. Taking into account the GW$_0$ electronic band gap, the computed binding energy is 0.34 and 0.66 for the singlet and triplet, respectively.  The experimental singlet binding energy has been reported several times for PPV and its derivatives and ranges from 0.06 to 1.0 eV~\cite{Moses2001,Alvarado1998,Campbell1996,Barth1997,Leng1994,Chandross1997}. The potential explanation of this spread lies in a complex morphology of the plastic films, which consists of small crystallites and amorphous domains~\cite{Masse1990}. Therefore, depending on the synthesis methods, certain samples can inherit electronic properties of either crystal or single chain, which are known to deviate significantly for polymers~\cite{Horst2000,Ruini2002,Horst2002,Tiago2004}.

\subsection{Fragment-based analysis}

\begin{figure}[tb]
\centering
 \includegraphics[width=\linewidth]{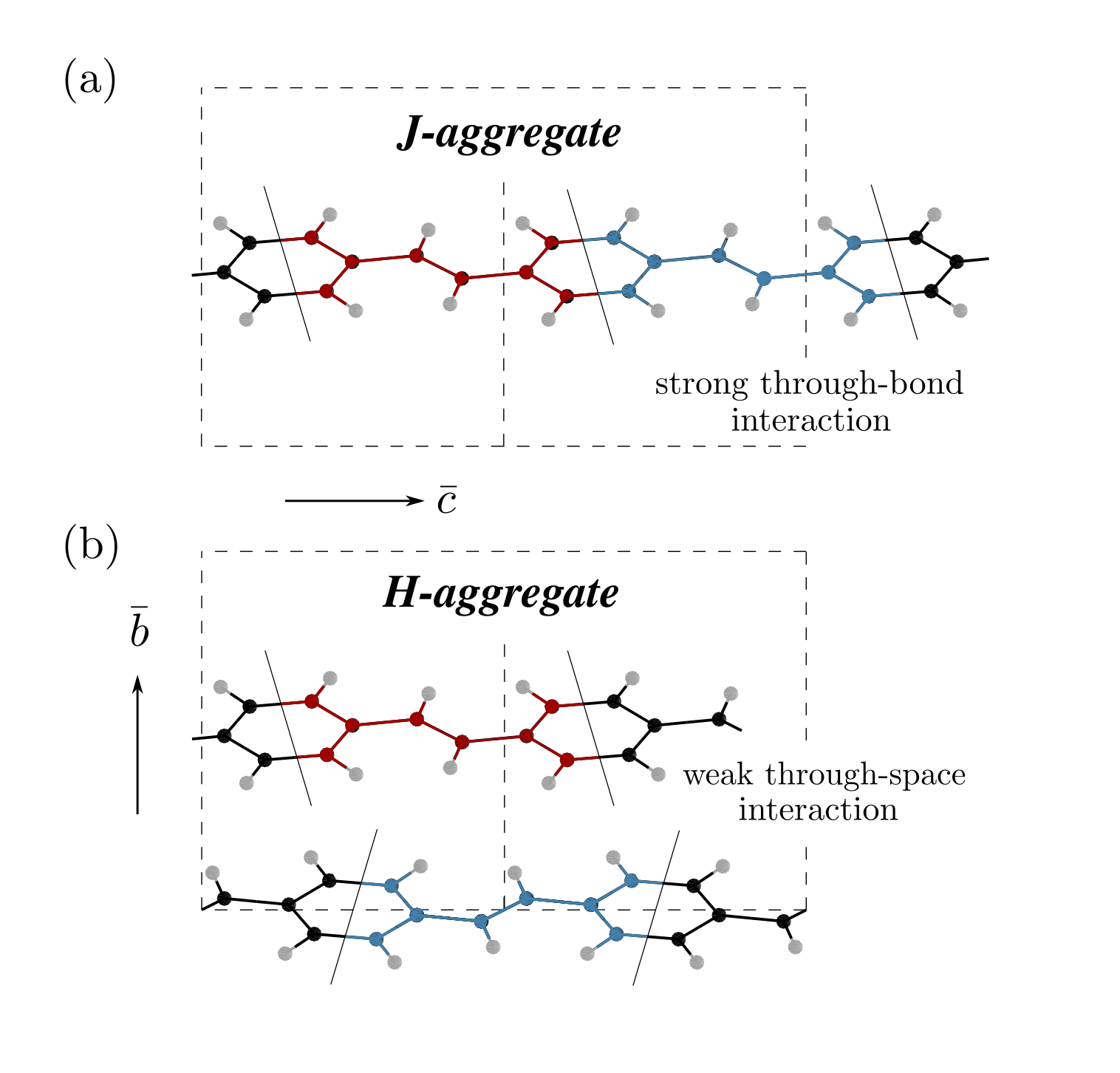}
 \caption{J-type (a) and H-type (b) aggregates in PPV crystal.}
 \label{fig:aggregates}
\end{figure}

In order to understand the static and dynamic properties of excitons in organic semiconductors, the fragment-based tight-binding (TB) model is often utilized. This approximation represents a solution of the multichromophoric systems of arbitrary dimension in terms of monomeric (on-site) energies $e_s$ and the nearest-neighbor transfer integral $J$. This is described by the Hamiltonian:
\begin{equation}
\label{eq:frenkel}    
  \hat{H}_\textrm{TB}=e_s \sum_m \hat{a}^\dag_m\hat{a}_m + J\sum_m(\hat{a}^\dag_m\hat{a}_{m+1} + \hat{a}^\dag_{m+1}\hat{a}_m)
\end{equation}
where the ladder operators act on local Frenkel excitons. Introduced as an empirical model, it requires a suitable parameterization against high-level electronic structure theory. To this end, we employed the density functional theory-based multireference configuration interaction (DFT/MRCI) method~\cite{Marian2019} in the R2016 formulation~\cite{Lyskov2016} with split-valence SV(P) atomic basis set~\cite{Schaefer1992} for all molecular calculations in the ensuing discussion. The solid state solvation effect was treated within the conductor-like screening model (\textsc{COSMO})~\cite{Klamt1993} in conjunction with the \textsc{TURBOMOLE}~\cite{TURBOMOLE} package. To evaluate the excited state energies we adopted the static value for the dielectric constant, $\epsilon=3$, mimicking the electrostatic screening due to adjacent PPV chains~\cite{GomesdaCosta1993,Horst2002}. Our approach, referred to as diabatization, relies on a pointwise mapping of the Hamiltonian eigenvalues in Equation~\ref{eq:frenkel} to the DFT/MRCI spectrum by numerically optimizing the TB parameters. In this way, we aim to compute the triplet-transfer integral for the J-dimer (two neighboring vinylene fragments on a single PPV chain, see Figure~\ref{fig:aggregates}-a) and for the H-dimer (two neighboring vinylene fragments on adjacent chains, see Figure~\ref{fig:aggregates}-b).

We begin by describing the excitonic structure of the single PPV chain, which resembles photophysical properties of J-aggregate~\cite{Spano2014}. For weakly coupled or spatially separated molecules in an aggregate, it is usually sufficient to consider two neighboring chromophores to determine the $e_s$ and $J$~\cite{Yost2012,Stehr2014,Spiegel2017}. That is, however, not the case for extended $\pi$-conjugated systems, where frontier HOMO and LUMO orbitals are spread across the chain. The effects of a conjugation break can be circumvented by adding two complementary fragments at the edges of the oligomer. Therefore, in order to find the interchromophore coupling in J-aggregate we truncated the PPV chain with optimized crystal parameters to the PPV$_4$ oligomer and analyze its electronically excited states in terms of the TB model. Table~S1 in supplemental information (SI) summarizes result of the DFT/MRCI calculations.

\begin{figure}[tb]
\centering
 \includegraphics[width=\linewidth]{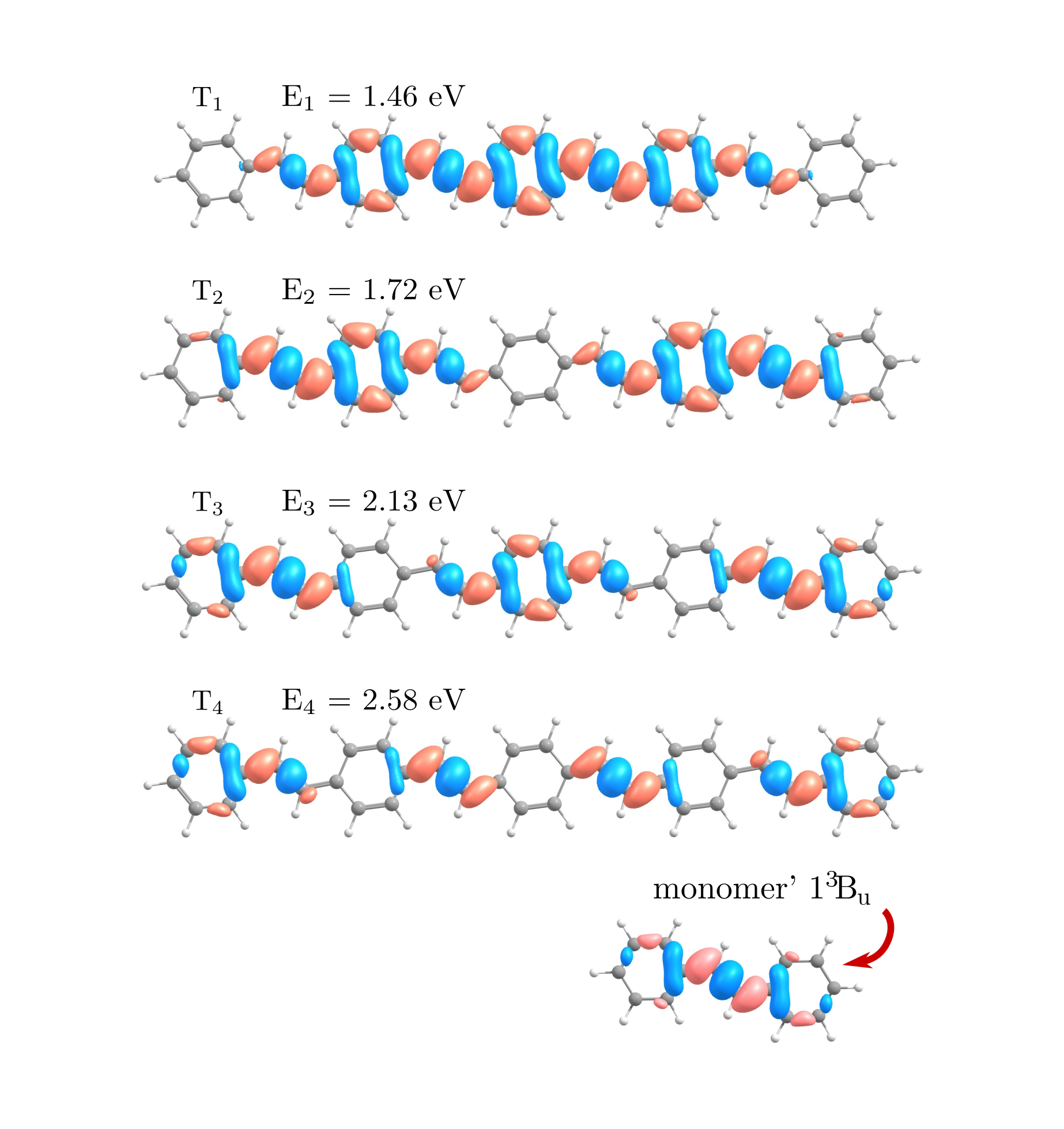}
  \caption{Electronic density difference of the triplet excited states with respect to the ground state in PPV$_4$ oligomer as computed with the DFT/MRCI method. An electronic density gain/loss is painted in red/blue.}
 \label{fig:densities}
\end{figure}

As previously reported the lowest singlet excited states of the PPV oligomers are formally given by a linear combination of Frenkel excitons residing on respective vinylene fragments~\cite{Ruini2002,Binder2013}. From the electronic structure point of view, they are built by promoting one electron from the HOMO to the LUMO orbitals yielding a set of localized 1$^1\!$B$_\textrm{u}$ states. By looking at the composition of the underlying wave functions, we observed that some of the low-lying singlet states exhibit substantial double excitation character. This is a distinctive fingerprint of the multiconfigurational 2$^1\!$A$_\textrm{g}$ state, which is a commonplace in the short-chain linear polyenes. It is expected that all excited-state properties particular to polyenes are transferable to PPV, such that the vibronic effects can facilitate ultrafast energy relaxation between local 1$^1\!$B$_\textrm{u}$ and 2$^1\!$A$_\textrm{g}$ states~\cite{Komainda2013,Komainda2016,Lyskov2017}. Consequently, the time-evolution of the singlet in PPV should incorporate (i) a description of two on-site excitons, (ii) the intrafragment vibronic coupling and (iii) interfragment hopping~\cite{Fornari2016}. However, the time-evolution of the singlet state in PPV is not the focus of this manuscript and will be addressed in a future study.

\begin{figure}[tb]
\centering
 \includegraphics[width=\linewidth]{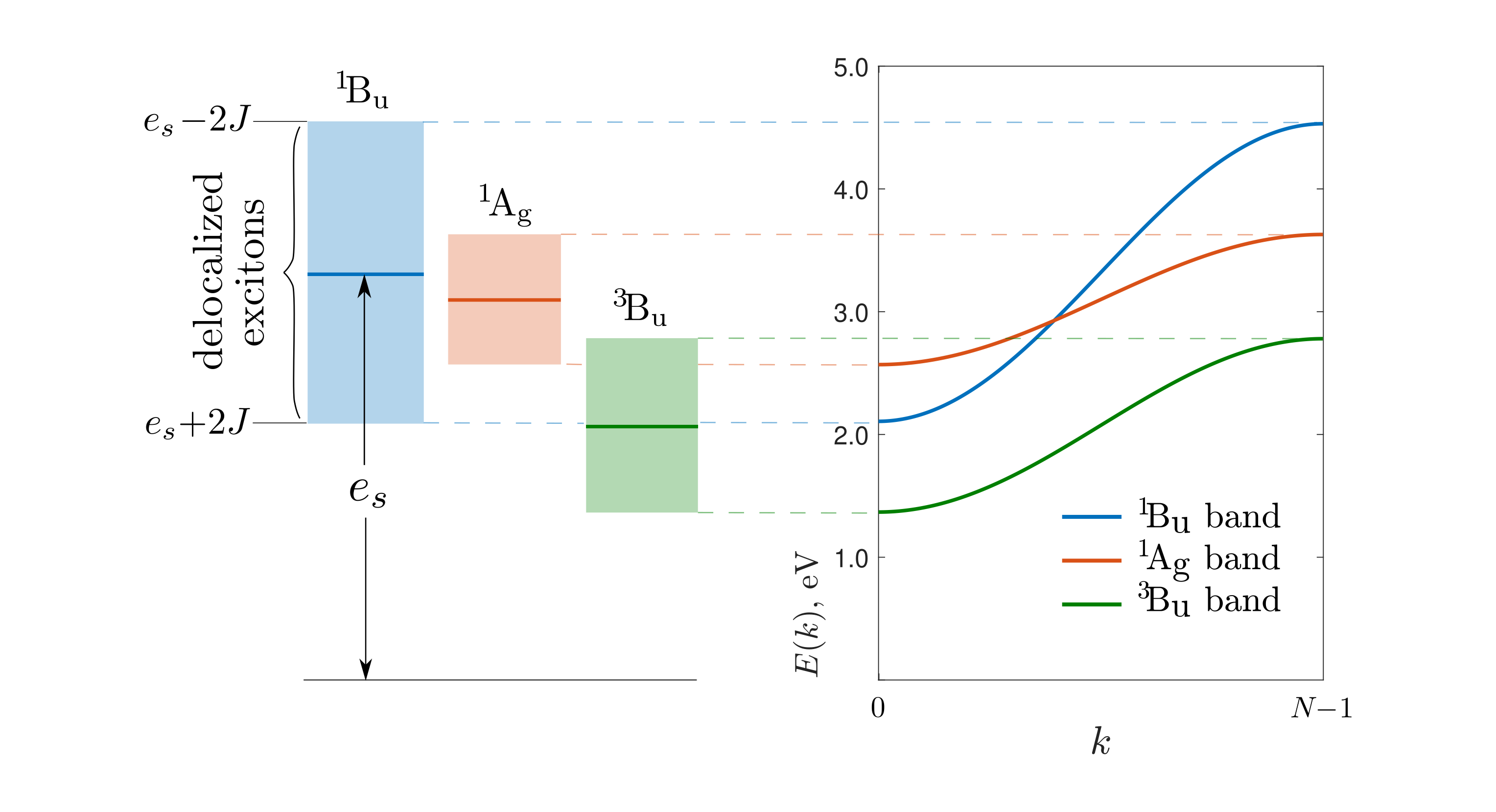}
  \caption{The singlet and triplet excitonic bands in PPV chain originating from the nearest-neighbor interaction approximation.}
 \label{fig:molbands}
\end{figure}

In turn, the four lowest triplet states in PPV$_4$ are of the $^3\!$B$_\textrm{u}$ character, see Figure~\ref{fig:densities}. This makes the mapping of the DFT/MRCI outcome to the TB Hamiltonian relatively straightforward, as each of the constituent monomers carries one exciton index associated with the Frenkel HOMO$\rightarrow$LUMO transition. Their linear combinations provide a set of standing excitonic waves delocalized over the size of the oligomer, which is similar to the {\it 'particle in the box'} model. The result of numerical optimization of the TB parameters for the lowest singlet and triplet states is shown in Table~\ref{tab:tb-parameters}. The knowledge of diagonal and off-diagonal couplings in Equation~\ref{eq:frenkel} immediately yield the supramolecular wave functions expanded in the basis of the local states $|m\ra$~\cite{Kuehn2011}:
\begin{equation}
  \label{eq:k-wf}
  |k\ra = \sqrt{\frac{2}{N+1}} \sum_{m=1}^{N} \sin\Big[{\frac{\pi m (k+1)}{N+1}}\Big] |m\ra
\end{equation}
with corresponding energies,
\begin{equation}
  \label{eq:k-e}
  E(k)=e_s+2J\cos \Big[\frac{\pi (k+1)}{N+1} \Big]
\end{equation}
where $k$ runs as $k=0,1,2,\ldots,N-1$, effectively reflecting the number of exciton density nodes. Equation~\ref{eq:k-e} indicates that for sufficiently large number of conjugated fragments ($N$) the on-chain excitonic band is formed with 4$J$ bandwidth~\cite{Brixner2017} (see Figure~\ref{fig:molbands}). We label the bands in Table~\ref{tab:tb-parameters} according to the symmetry notation of the respective monomeric components. The energy and dispersion relation of excitons in a band are defined by the TB parameterization. For example, the on-site 1\singbu state is energetically higher than the 2\singag state, being 3.32 eV and 3.10 eV, respectively. However, when a superposition is formed, the totally symmetric combination, i.e. $|k=0\ra$, of the 1\singbu states at 2.11 eV is energetically stabilized to a larger extent compared to the \singag counterpart, which is at 2.57 eV. This is achieved due to strong interchromophore coupling between two dipole-allowed states in J-type aggregate. The electronic $|k=0\ra$ state at 1.37 eV represents the steady-state solution of the time-evolving triplet exciton on the polymer chain.

\begin{table}[tb]
\begin{tabular*}{0.48\textwidth}{@{\extracolsep{\fill}}l ccc}
\hline\hline\\[-0.9em]
& \singag & \singbu & \tripbu \\[0.2em]
\hline\\[-0.9em]
$e_s$, (eV) & 3.099  & 3.319  & 2.074 \\
$J$, (eV)   & -0.265 & -0.606 & -0.353   \\
$E$, (eV)   & 2.57   & 2.11   & 1.37     \\[0.2em]
\hline\hline
\end{tabular*}
\caption{Optimized on-site energy ($e_s$), hopping integral ($J$) and excitonic band energy ($E$) in PPV chain of infinite length}
\label{tab:tb-parameters}
\end{table}

Using the PPV$_4$ oligomer we defined the on-site triplet energy (2.07 eV) and through-bond coupling between sites ($-0.35$ eV). In a similar way the interchain hopping integral can be extracted, assuming non-zero interaction for nearest neighbors only. We treat two PPV$_3$ oligomers packed in herringbone assembly as in the crystal structure. For the sake of consistency, the first off-diagonal elements of the TB matrix were set to those optimized for individual oligomers, and true off-diagonals were relaxed to reproduce the DFT/MRCI energies (see SI). The resulting through-space triplet coupling $J$ between inner sites of oligomers was found to be 0.0114 eV, that is much lower than the strength of $\pi$-conjugation by at least order of magnitude. Therefore, different regimes of triplet transport along different crystal axes of the PPV can be anticipated.

\subsection{On-site potentials and interfragment coupling}

Depending on the aggregation type, transformations of the ground state geometry can hamper or facilitate exciton propagation. Despite the fact that static disorder can be neglected in the ideal crystal, the transport occurs naturally across the energy landscape, which is spawned by thermal fluctuations of nuclear coordinates. Moreover, strong exciton-phonon coupling can be ascribed to another source of dynamic disorder, which rearranges local nuclear coordinates when an exciton hits a particular fragment. The structural deformation of the polymer geometry can be qualitatively reflected by detuning the on-site energy and changing the excitonic coupling in Equation~\ref{eq:frenkel}. This requires the explicit inclusion of the vibrational modes into the Hamiltonian of Equation~\ref{eq:frenkel}, which only has terms representing excitonic interaction.

\begin{figure}[tb]
\centering
 \includegraphics[width=\linewidth]{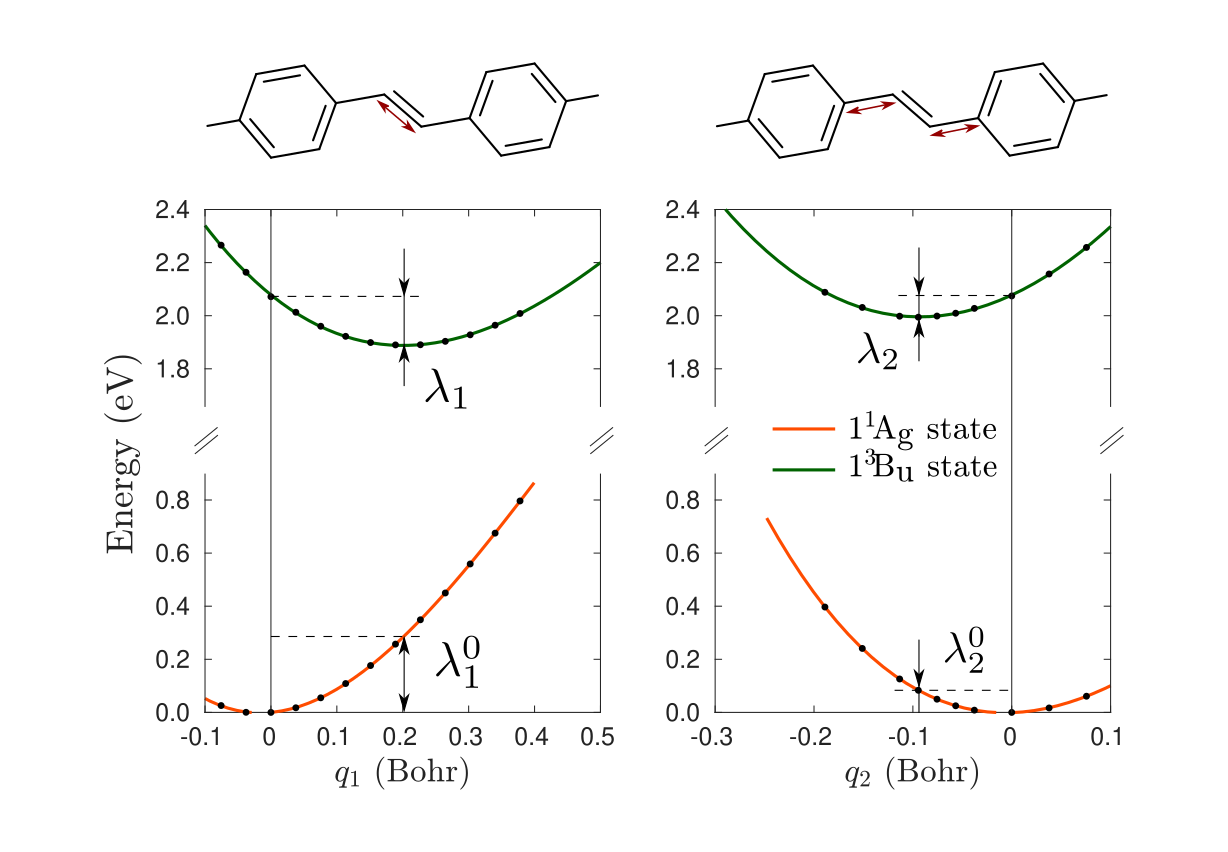}
  \caption{On-site PES along the C--C bond-length alternation coordinates. The nuclear reorganization energies are $\lambda_1/\lambda_1^0=190/283$ meV and $\lambda_2/\lambda^0_2=93/61$ meV.}
 \label{fig:potentials}
\end{figure}

In absence of static disorder all fragments of an infinite PPV chain behave in the same fashion upon local distortion of the nuclear framework. We analyzed the DFT/MRCI solution for PPV$_4$ oligomer by displacing the geometry of two inner sites while the terminal sites are remained unchanged. The on-site potential energy surfaces (PES) were computed by solving the reverse eigenvalue problem at every geometry point as discussed in details in Ref.~\citenum{Binder2014}. In the present case, a third order polynomial function was chosen to fit the electronic structure data. To describe the nuclear degrees of freedom (DOF), we used internal curvilinear coordinates which characterize the most important vibrations coupled to the local 1\tripbu state. They comprise stretching of nominally single bonds ($q_2$), and elongation of the double bond ($q_1$) on vinylene as shown in Figure~\ref{fig:potentials}. By the diabatization condition these DOFs do not affect the inter-site coupling, but solely change the on-site energy. The energy modulation of 0.19 eV along $q_1$ and 0.09 eV along $q_2$ reflects the nuclear response to the triplet exciton, in that C--C bonds become nearly equal at the minimum of the 1\tripbu PES. Although the vibrational DOFs are not generally separable for the molecular aggregate, the bond-length alternation on neighboring fragments can be grouped in pairs in a symmetric ($Q_1$ and $Q_3$) and antisymmetric ($Q_2$ and $Q_4$) way, generating a new set of common nuclear coordinates. In the following discussion we call them on-site modes.

Another important class of nuclear DOFs to consider are the so-called coupling modes. Unlike the on-site modes they modify the coupling strength between two interacting triplets. We analyzed this modification when the bond length on a junction ($Q_5$), i.e. on the phenylene segment encapsulated between two Frenkel sites, was changed. We refer a reader to the supplemental information for the definition of nuclear DOFs and DFT/MRCI PESs. Although the on-site energy is not changed significantly, the excitonic coupling along this coordinate portrays a steeply descending potential-like linear function. Such an active modulation of $J$ with $-1.55$ eV/Bohr slope provides an evidence for strong vibronic interaction in PPV. Besides the bond-length alternation on phenylene fragment, we consider the torsions about the C--C bond connecting the phenylene and vinylene moieties ($Q_6$ and $Q_7$). At its core these modes reduce the degree of $\pi$-conjugation between neighboring nearly co-planar Frenkel sites leading to one of the dominant conformational defects in polymers at large twisting amplitude~\cite{Tozer2012,Binder2013}. This set of 7 nuclear DOFs for the J-dimer allows compact characterization of on-chain exciton relaxation in accordance with previous computational study of PPV oligomers~\cite{Panda2013,Binder2013}.

\begin{figure}[tb]
\centering
 \includegraphics[width=\linewidth]{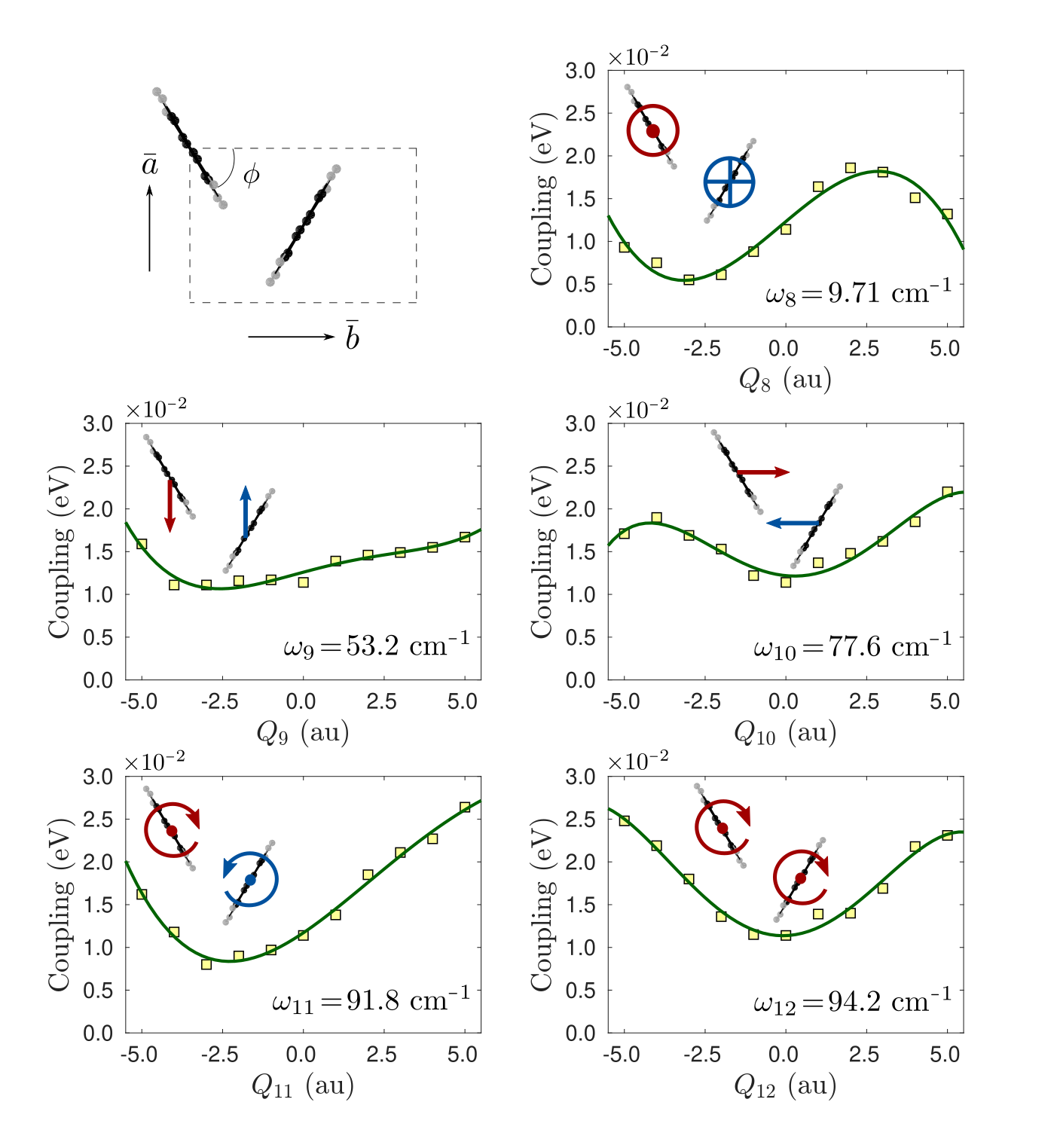}
  \caption{Modulation of the through-space Dexter coupling resulting from the exchange interaction between localized electrons. Horizontal axes describe geometry displacement in mass-weighted normal coordinates.}
 \label{fig:pi_overlap}
\end{figure}

Obviously, the on-site modes ($Q_1$, $Q_2$, $Q_3$ and $Q_4$) have to be taken into account for the H-dimer, as well. In this case two monomers are not linked by $\pi$-conjugation, and thus the coupling modes have to be revised in regards to their effect on excitonic coupling. For spatially separated monomers the triplet-triplet coupling is mediated by electronic exchange interaction~\cite{Ostroverkhova2016}. At long distances the exchange is known to decay exponentially with interparticle radius, and, at shorter distances, it depends strongly on the {\it 'physical'} overlap between localized electronic wavefunctions, i.e. on mutual orientation of the monomers. In order to understand how change of spatial orientation of adjacent chains modifies the magnitude of Dexter coupling, we analyzed the low-frequency modes of the phonon spectrum sampled at the $\Gamma$ point of the crystal. We performed the DFT/MRCI calculations of two PPV$_3$ oligomers from adjacent chains and performed diabatization, as described in the supplemental information, to evaluate how interchain modes modify excitonic coupling. These modes can be viewed as slippage of one chain with respect to another along different crystal axes ($Q_8$, $Q_9$ and $Q_{10}$), and as in-phase and counter-phase rotations ($Q_{11}$ and $Q_{12}$) about intrinsic axis of individual chains in Figure~\ref{fig:pi_overlap}. The aforementioned DOFs do not alter internal bond-angle coordinates in PPV chains, but modify the mutual orientation of oligomers in space, thereby changing the intersite coupling in H-aggregate arrangement. Displacement of the Frank-Condon (FC) geometry along these coordinates can be kinetically activated by ambient temperature. Therefore, we account for 9 nuclear coordinates for the H-dimer.

Using one-dimensional potentials discussed above we built the multidimensional PES for the localized electronic states. Because the computed triplet-triplet coupling at the FC point is different by orders of magnitude for the J- and H-aggregate arrangement, this allowed us to process intrachain and interchain transport components independently.

\section{Mobility of triplet exciton}

In the quantum domain there are two constructive ways to elucidate nonequilibrium phenomena. The first approach is based on a time evolution of a combined system-environment which usually involves solving a fairly complex Schr\"odinger equation. The dynamics for the central electronic system are often found by tracing out all interacting components of the environment~\cite{Binder2013,Binder2018}. The second approach relies on tracing out the environmental DOFs before solving the complex system under nonequilibrium settings. This often involves a certain degree of phenomenology in order to derive a reduced equation of motion for the system density operator~\cite{Cao2013,Mannouch2018}, known as a master equation. We believe that the best way to elaborate on the space-time evolving exciton in a material is to successively engage both approaches, as the transient Schr\"odinger dynamics between two equivalent segments can be extrapolated to an arbitrary number of aggregated molecules. Below we outline the coupled electron-nuclear Schr\"odinger equation and, in the following, discuss and interrelate its solution with the master equation approach.

\subsection{Vibronic Hamiltonian}

In the localized Frenkel basis four supramolecular electronic states can be drawn. We label them as $|10\ra$ and $|01\ra$ when the exciton is situated on one or another moiety of a dimer, the $|00\rangle$ for the ground state and $|11\ra$ for excimer configuration. The latter two are neglected in the present model, which yields the vibronic Hamiltonian for the two-level system:
\begin{equation}
  \label{eq:mctdh_ham}
  \hat{\bf H}_\textrm{vib} = \frac{1}{2} \sum_{i,j}  \frac{\textrm{G}_{ij}}{\partial Q_i \partial Q_j} \cdot \hat{{\bf 1}}_\textrm{el} + \hat{{\bf 1}}_\textrm{vib} \cdot
  \begin{bmatrix}
    e_s & J \\
    J & e_s \\
    \end{bmatrix}
  +
  \sum_i
\begin{bmatrix}
     u_{11} &  u_{12} \\
     u_{12} &  u_{22} \\
\end{bmatrix}
(Q_i)
\end{equation}
The first term describes the nuclear kinetic energy operator as formulated in terms of the $\textrm{G}$-matrix~\cite{Wilson1955}, the elements of which are tabulated in Ref.~\citenum{Frederick1999}. The second and third terms represent the coupling matrix at the FC point and the multidimensional PES, respectively. The multidimensional PES was constructed using a linear vibronic coupling model~\cite{Koppel1984}. Because the $|10\ra$ and $|01\ra$ states are composite, the diagonal elements of the PES matrix are defined through the sum of the one-dimensional on-site 1\tripbu and 1\singag potentials ($v_\textrm{ex}$ and $v_\textrm{gs}$). In the following equation the superscript designates the site number, so that the $|10\ra$ PES reads 
\begin{equation}
  \label{eq:diag2}
   u_{11}(Q_i) =  v_\textrm{ex}^{1} (Q_i) + v_\textrm{gs}^{2} (Q_i) \\
\end{equation}
and {\it vice versa} for the $|01\ra$. The off-diagonal elements combine all coupling modes. We recall that the change of on-site energy depends only on nuclear coordinates, which modifies the geometry of a particular site according to the employed diabatization scheme.

In order to describe the time evolution of the system with $f$ nuclear DOFs, we adopt the multiconfigurational wave function:
\begin{equation}
  \label{eq:mctdh_wf}
  \Psi(Q_1...Q_f,t) = \sum_{j_1}^{n_1} \ldots \sum_{j_f}^{n_f} A_{j_1\ldots j_f}(t) \prod^f_{k=1} \phi^{(k)}_{j_k} (Q_k,t)
\end{equation}
where $A_{j_1\ldots j_f}$ are time-dependent expansion coefficients and $\phi^{(k)}_{j_k}$ are expansion functions of each oscillator $Q_k$. The spin-free vibronic Hamiltonian $\hat{\bf H}_\textrm{vib}$ in Equation~\ref{eq:mctdh_ham} and the vibrational wavefunction in Equation~\ref{eq:mctdh_wf} enter the Schr\"odinger equation: 
\begin{equation}
  \label{eq:shroedinger}
  i\hbar \frac{\partial}{\partial t} \Psi = \hat{ \bf H}_\textrm{vib} \Psi
\end{equation}
for variational optimization of the wavefunction parameters at every time interval of the propagation. Thus, we treat the correlated exciton-nuclear dynamic problem within the fully quantum time-dependent scheme using the Heidelberg \textsc{MCTDH} package~\cite{MCTDH}. See supplemental information for details of the MCTDH basis.

We carried out 300 relaxation jobs at finite temperature (300 K) to obtain a thermal sampling of the vibrational wavefunction on the ground electronic state $|00\ra$ as prescribed in the stochastic random-phase wavefunction approach~\cite{Nest2007}. Then, each of the wavepackets was energetically lifted to the $|10\ra$ potential to mimic an exciton injection. Subsequently, the wavepackets evolve in time on two coupled seven-dimensional PESs for the J-aggregate and two nine-dimensional PESs for the H-aggregate. The results of the wavepacket propagation discussed below describe a statistical average over all initial random-phase realizations.

\subsection{Intrachain dynamics}

\begin{figure}[tb]
  \centering
   \includegraphics[width=\linewidth]{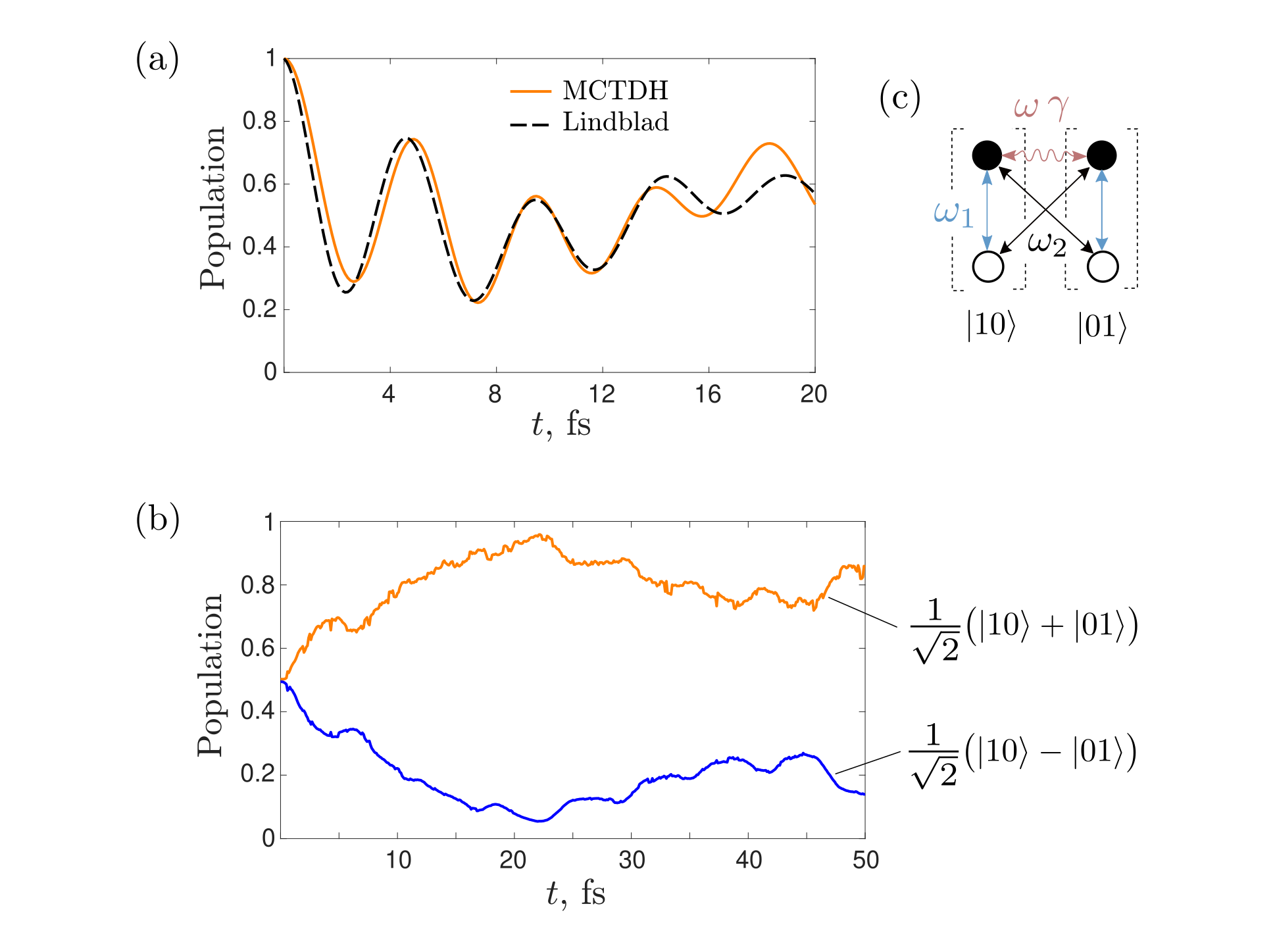}
  \caption{(a) The population dynamics of triplet exciton between two J-aggregated sites in local (nonadiabatic) basis. Orange curve displays the {\it ab-initio} results and dashed black curve is a solution of the master equation. (b) The population dynamics of triplet exciton between two J-aggregated sites in mixed (adiabatic) basis. (c) Schematic representation of the master-equation model for the J-dimer. Black and white circles denote physical and auxiliary states, respectively.}
  \label{fig:two_fragments}
  \end{figure}

For the case of J-aggregation the on-chain energy transfer can be characterized by two intimately related processes: (i) the site-to-site energy migration which proceeds in a wave-like regime; (ii) formation of coherence properties when on-site excitons superpose locally. In the following, the $|10\ra$ and $|01\ra$ states designate an exciton residing on one or another moiety of the J-dimer. Figure~\ref{fig:two_fragments}-a shows the temporal evolution of a triplet exciton after populating the $|10\ra$ state. Due to strong through-bond coupling, the initial population transfer occurs in a form of Rabi oscillations. The rapid oscillations are accompanied by structural deformation of the backbone chain, which introduces a partial attenuation of the probability flopping over short times. Upon activation of the on-site bond-length alternation modes, the signal's centroid shows a bias towards the energy accepting $|01\ra$ state at $t=10$ fs, corresponding to half of the C--C vibrational period. The characteristic trend of ultra-fast oscillations following carrier frequency of localized vibrations persists at longer timescale and can be thought of as an exciton-polaron~\cite{Binder2013,Binder2018}.

Besides ultrafast population redistribution in the J-dimer, the coherence between two Frenkel sites develops rapidly after initial exciton generation as shown in Figure S4 in SI. Although the solution of the Schr\"odinger equation is given in the on-site basis (nonadiabatic), which is in the spirit of MCTDH, the underlying PESs alongside the time-dependent MCTDH wavefunction can be transformed into the adiabatic representation. This yields the system dynamics in a purely entangled basis of $1/\sqrt{2}(|10\ra\pm|01\ra)$  states, as shown in Figure~\ref{fig:two_fragments}-c. As one can see, two spatially-separated Frenkel excitons interfere within the characteristic time of the high-frequency C--C mode of $\sim$20 fs. The same time interval for coherence transfer was detected in linearly aligned MEH-PPV in a two-time anisotropy decay experiment~\cite{Collini2009}. From this we conclude that the local nuclear deformations introduce the site-to-site coherence. Due to the initial conditions in our simulations, some amount of energy excess exists in the system leading to admixture of the second excited state. The energy excess can be drained out by an external bath which deactivates the bond-length alternation oscillations, producing a relaxed vibrational state. On a longer timescale, energy dissipation through a non-zero system-bath interaction causes exciton self-localization~\cite{Cao2013,Consani2015,Mannouch2018}. Nonetheless, even in the case of a vibrationally hot exciton, the cooperative motion of electrons and nucleus governs coherent wavefunction delocalization along the conjugated backbone.

The unitary evolution of excitons in a polymer chain of higher dimension necessitates including additional nuclear DOF at each site. This inevitably increases the computational demands and makes the numerical treatment of the system dynamics unfeasible. Here, we circumvent numerical intractibility for large PPV chains using a master equation approach. Using the Schr\"odinger picture for the two-state model we derive key parameters characterizing the transient exciton evolution, which are then fed into a master equation with far fewer degrees of freedom. In its most common Lindblad form, the master equation describes time evolution of an electronic system (excitons) weakly coupled to the environment (vibrational bath). However, certain modifications are required in order to account for strong exciton-phonon coupling. Therefore, in addition to the environment we introduce auxiliary electronic states in the master equation to mimic the polaronic properties of an exciton. Because the chosen set of nuclear DOF was preliminarily split into two groups of different functionality, i.e. the on-site and coupling modes, they formally induce different types of exciton relaxation. The on-site vibrations introduce oscillatory transition of the localized excitons, coupled by $\omega$, from energy resonance at the FC point to the energy-detuned region (when a local exciton is energetically stabilized due to reorganization energy of the local potentials). This time-dependent energy detuning can be captured by adding auxiliary electronic states in the effective system Hamiltonian $\hat{H}_\textrm{eff}$, each of which is linked to the individual electronic DOF by coupling $\omega_1$ and interacts with the neighboring physical states with $\omega_2$ (see  Figure~\ref{fig:two_fragments}-c). In turn, the coupling modes stimulate phenomenological dephasing and damps ultrafast Rabi oscillations. We contract them to one effective environmental potential acting locally on pairs of interacting monomers with strength $\gamma$. Tracing out this environment leads to the master equation for the system density matrix $\rho$:
\begin{equation}
  \label{eq:lindblad}
  \dot{\rho} = -\frac{i}{\hbar}[\hat{H}_\textrm{eff},\rho] + \sum_i \gamma_i \Big( \hat{L}_i \rho \hat{L}_i^\dag - \frac{1}{2}\hat{L}_i^\dag \hat{L}_i\rho - \frac{1}{2}\rho \hat{L}_i^\dag \hat{L}_i \Big)
  \end{equation}
with the summation running over interacting pairs. The Lindblad jump operators are Pauli $\sigma$-matrices, i.e. $\hat{L}=\hat{\sigma}_x$. Equation~\ref{eq:lindblad} requires four parameters, three system and one environmental parameter, which we optimize with respect to the MCTDH solution.

\begin{figure}[tb]
  \centering
   \includegraphics[width=\linewidth]{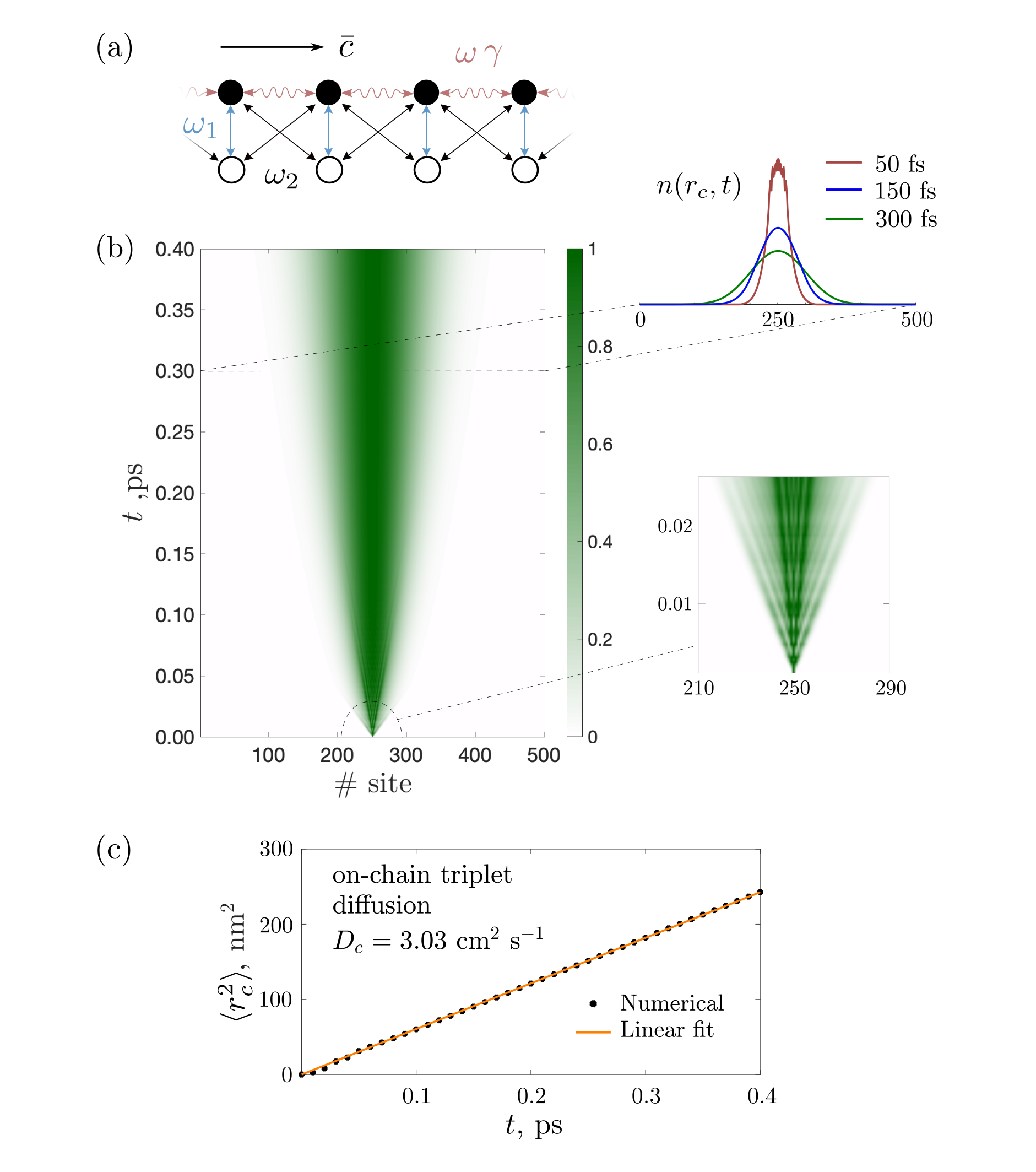}
  \caption{(a) Graphical representation of the master-equation model for extended PPV oligomer. (b) Time evolution of the triplet exciton density along the PPV chain computed using the master-equation approach. (c) The mean square displacement of on-chain triplet exciton.}
  \label{fig:onchain}
  \end{figure}

In reality the exciton density is not artificially confined within the length of two fragments but continuously propagates towards ends of the chain. We characterize the on-chain transport by the main features of the short-time interfragment dynamics spanning a time period from 0 to 20 fs. The dashed black line in Figure~\ref{fig:onchain}-a shows triplet population of the $|10\ra$ state obtained by pre-parameterized Lindblad equation (see Table~S3 in SI). The resulting population dynamics imitates essential features of the Schr\"odinger evolution: high-frequency fluctuations as well as  Rabi-type oscillations with gradual attenuation. The set of optimized Lindblad constants yields an efficient parameterization of the triplet density evolution in J-type dimer, and can be applied for the arbitrary size PPV chain with equally linked monomers, as depicted in Figure~\ref{fig:onchain}-a.

To demonstrate on-chain evolution of triplet exciton we numerically investigated a system containing 501 linearly aligned units using our master equation method. The initial density was placed on the central site of the oligomer to allow balanced migration in both directions. Figure~\ref{fig:onchain}-b shows results of our simulations where two different transport components can be observed. On the short time scale, when rapid oscillation dominates over $\gamma$, the exciton shows ballistic expansion as a result of the initial confinement within the size of one monomer. At longer times the transport proceeds in a diffusive regime. In this case the homogeneous transport without external current sources can be described by a diffusion equation:
\begin{equation}
  \label{eq:diffusion}
  \frac{\partial  }{\partial t} n(r_c,t) = D_c \nabla^2 n(r_c,t)
  \end{equation}
where $D_c$ is on-chain diffusion coefficient. The solution of Equation~\ref{eq:diffusion} using the initial condition from the master-equation simulations is given by Gaussian profile:
\begin{equation}
  \label{eq:dif_solution}
  n(r_c,t)= \sqrt{ \frac{1}{4\pi D_ct}} \exp\Big( -\frac{r_c^2}{4D_ct}\Big)
  \end{equation}
which characterizes the probability distribution of a space-time evolving exciton density with fixed center of mass. The diffusion constant $D_c$ is linearly related to the position variance as $\la r_c^2\ra=2D_ct$. The distance the exciton travels when undergoing a hopping to the neighboring site is taken to be length of the vector connecting the gravity centers of vinylene moieties. Linear regression of position variance of on-chain exciton returns a value of $D_c=3.03$ cm$^2$s$^{-1}$, see Figure~\ref{fig:onchain}-c.

\subsection{Interchain dynamics}

The high space symmetry group of the PPV crystal stipulates for identical interaction between a particular chain and all surrounding chains. Because the Dexter coupling between two sites in H-aggregate chains is relatively weak compared to those in J-aggregate, we can ignore high-frequency oscillations within the so-called secular approximation. It essentially assumes that within the nearest-neighbor interaction picture the chain-to-chain hopping of exciton density is qualitatively the same as the dynamics of site-to-site hopping.  Following our strategy, we solve the coupled exciton-vibrational dynamics using the Schr\"odinger equation for two H-aggregated monomers, and parameterize the master-equation for triplet density transfer in nanosized PPV crystal. In this case the two-state vibronic Hamiltonian comprises four on-site modes which are built from on-site PES (Figure~\ref{fig:potentials}) and five low-energy coupling modes (Figure~\ref{fig:pi_overlap}) as defined in previous sections.

\begin{figure}[tb]
\centering
 \includegraphics[width=\linewidth]{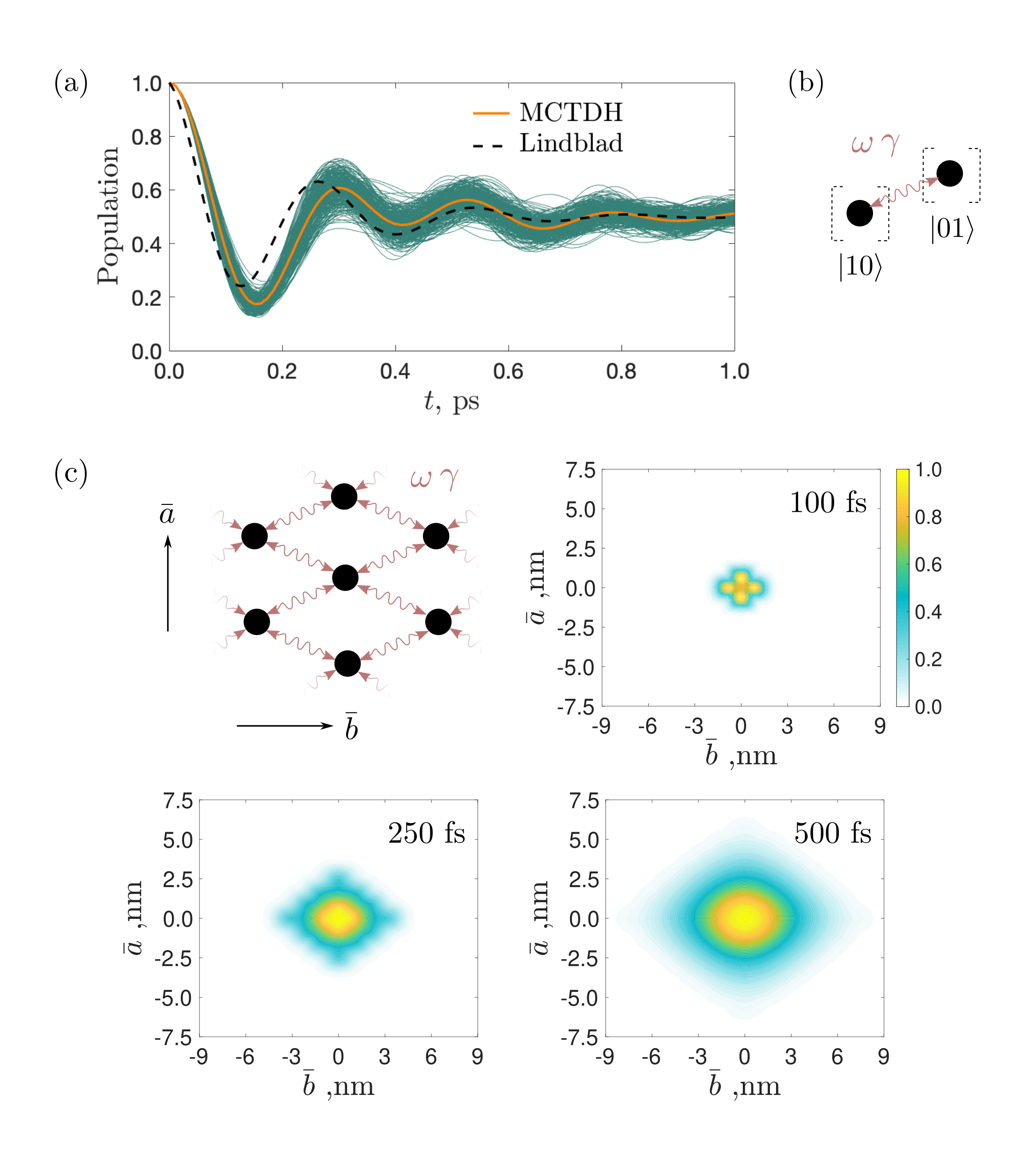}
  \caption{(a) Time evolution of triplet exciton migrating between two H-aggregated sites. Green curves show the propagation result of initial random-phase wavefunctions; orange curve shows arithmetic averaging over stochastic realizations; black stripes is optimized Lindblad solution. (b) Schematic representation of the master-equation model for the H-dimer. (c) The master-equation model for an exciton dispersing in the $\bar{a} \bar{b}$ plane and space-time evolving triplet density between adjacent PPV chains of the crystal.}
 \label{fig:interchain}
\end{figure}

Figure~\ref{fig:interchain}-a displays an exciton evolution in the H-dimer at an ambient temperature of 300 K computed by the MCTDH method, where the $|10\ra$ and $|01\ra$ states denote an exciton residing on one or another monomer. Notably, the interchain dynamics occurs at different timescale compared to the intrachain case. The MCTDH curve is described by Rabi oscillations, having period of $\sim$250 fs, with gradual attenuation of oscillatory amplitude. The population of the $|01\ra$ and $|10\ra$ states is equilibrated within 1 ps, during which no interchain coherence develops (Figure~S4 in SI). In parameterization of the master-equation model we omitted the interaction of exciton density with on-chain modes, as the main dynamical trend is clearly drawn by the low-frequency coupling modes. Similar to the intrachain case, we combine coupling modes into one effective environment $\gamma$ acting on electronic DOFs. This yields two Lindblad parameters for triplet dynamics in the H-dimer, i.e. interaction $\omega$ and environment $\gamma$ (Figure~\ref{fig:interchain}-b), which were optimized to fit results of the Schr\"odinger equation.

To model interchain transport we built a [001] slab containing 25 repeating unit cells along the $\bar{a}$ axis and 25 cells along the $\bar{b}$ axis of the crystal. The initial exciton was placed on the innermost site and subsequently propagated in a two-dimensional matrix for 1 ps. Figure~\ref{fig:interchain}-c shows femtosecond snapshots of triplet density migrating across the PPV chains, which is mediated solely by the herringbone neighbors. Similar to the exciton migration along the chain, the ballistic expansion of triplet density is observed at initial time.
Within characteristic timescale of $1/\gamma\sim600$ fs the transport transitions to the diffusive regime with the exciton qualitatively behaving as a classical Brownian particle. The triplet energy is progressively distributed over the $\bar{a}\bar{b}$ plane in a form of two-dimensional Gaussian function:
\begin{equation}
  \label{eq:dif2d}
n(r_a,r_b,t) = \frac{1}{4\pi t}\sqrt{\frac{1}{D_aD_b}}  \exp \Big(-\frac{r_a^2}{4D_at}-\frac{r_b^2}{4D_bt} \Big)
\end{equation}
Thus, the probability distribution of the dispersive ellipsoid is regulated by two constants along respective crystal axes $D_a$ and $D_b$. We fit the numerical results on Figure~\ref{fig:interchain}-b to the Equation~\ref{eq:dif2d} and find the interchain diffusion constants for triplet exciton of  $D_a=2.48\cdot10^{-2}$ cm$^2$s$^{-1}$ and $D_b=4.18\cdot10^{-2}$ cm$^2$s$^{-1}$.

\subsection{Final remarks}

The computed diffusion coefficient along the PPV backbone $D_c=3.03$ cm$^2$s$^{-1}$ stands out compared to molecular semiconductors, where the diffusion constant is typically much lower. This is due to the fact that intrachain excitonic domains are enclosed in one $\pi$-conjugation, where the strength of the electrostatic interaction between fragments is orders of magnitude higher than in organic solids with spatially separated molecular constituents. As a result, an exciton delocalizes coherently in a wave-like regime with retained phase information between the monomer wavefunctions. Such a high mobility is not surprising, as similar values for coherent exciton transport have been reported: for example, $D=1.9$ cm$^2$s$^{-1}$ was obtained using a vibronic Hamiltonian approach in organic DCVSN5 crystal~\cite{Arago2016}, and $D=32$ cm$^2$s$^{-1}$ was obtained from Monte-Carlo simulations in supramolecular light-harvesting nanotubes, which found an agreement with experiment~\cite{Caram2016}. In this regard intrachain diffusion can be compared with conventional inorganic semiconductors (for example Ref.~\citenum{Yuan2017,Cadiz2018}) where an exciton evolves in a band-like transport regime.

As expected, the computed components of interchain diffusion $D_a=2.48\cdot10^{-2}$ cm$^2$s$^{-1}$ and $D_b=4.18\cdot10^{-2}$ cm$^2$s$^{-1}$ are significantly slower than those along the fast axis. This is comparable to polyacene crystals~\cite{Williams1966,Akselrod2014,Najafov2010} where triplet transport is also mediated by through-space Dexter mechanism. We also note that the typical exciton diffusion coefficient in amorphous polymers of PPV family is lower by orders of magnitude than our values (see, for example, Ref.~\citenum{Markov2005,Mikhnenko2015}). This discrepancy stems from the large degree of conformational disorder, dislocation and coiling of elastic chains which greatly suppresses exciton migration in such solids.

We find that $\pi$-electron conjugation facilitates rapid long-range exciton migration in PPV and its chemical derivatives. A solid-state matrix which can carry triplets over a long distance is of particular importance in the design of efficient energy upconverters. Our results suggest that polymeric materials  with an extensive rigid backbone and long-range crystalline order provide high mobility and, hence, a high probability of triplets colliding and undergoing fusion before deactivating to the ground state.

\section{Conclusion}
  
In this paper we have investigated triplet energy transport in disorder-free crystalline PPV. We  determined electronic structure properties of triplet excitons by an effective mapping of the DFT/MRCI spectum of short-size oligomers to the Frenkel tight-binding model. We found that the local 1\tripbu electronic transition on vinylene fragments parameterizes the lowest triplet excitonic band in PPV. We obtained a fully quantum description of population transfer between two neighboring Frenkel sites in the J- and H-aggregate configuration, which is exclusively driven by electronic and vibronic coupling. Further, we established a connection between resulting Schr\"odinger dynamics and the master-equation approach, which allows us to study exciton propagation in extended systems in the context of cost-efficient Lindblad model. We obtained transient intrachain and interchain exciton density evolution and extracted respective triplet diffusion coefficients along the principal axes of the crystal. 

As expected, the triplet mobility is highly anisotropic with respect to the direction of exciton propagation. We found that triplet transport in PPV is characterized by  two essential components with remarkably different diffusion constants: fast intrachain, and slow interchain. Even in case of migration along the $\pi$-conjugated backbone, transport can be characterized by normal diffusion following the ultrafast wave-like ballistic spread of the initial nonequilibrium density. The diffusion constant along the $\bar{c}$ crystal axis was found to be $D_c=3.03$ cm$^2$s$^{-1}$. Electronic coherence between two neighboring Frenkel sites in J-aggregates develops within 20 fs and is stimulated by structural deformations of the carbon framework. No coherence between chains was observed. At room temperature (300 K) the corresponding diffusion coefficient along $\bar{a}$ and $\bar{b}$ axes are $D_a=2.48\cdot10^{-2}$ cm$^2$s$^{-1}$ and $D_b=4.18\cdot10^{-2}$ cm$^2$s$^{-1}$. We emphasize importance of the long-range order in solid-state polymeric material to exciton transport.

\section{acknowledgement}
This work was supported by the Australian Government through the Australian Research Council (ARC) under the Centre of Excellence scheme (project number CE170100026 and CE170100012). It was also supported by computational resources provided by the Australian Government through the National Computational Infrastructure National Facility and the Pawsey Supercomputer Centre. E.T. acknowledges support by the Deutsche Forschungsgemeinschaft through collaborative research center SFB 953.
%\end{acknowledgement}

\section{Supplementary Information}
The following files are available free of charge.
\begin{itemize}
\item ppv-si.pdf: the supplementary information file with DFT-PBE electronic band structure, DFT/MRCI excitation energies, definition of nuclear coordinates, details of the MCTDH basis, optimized Lindblad parameters.
\end{itemize}
%\end{suppinfo}
%% and also names the section correctlytly.
%%%%%%%%%%%%%%%%%%%%%%%%%%%%%%%%%%%%%%%%%%%%%%%%%%%%%%%%%%%%%%%%%%%%%

%\bibliography{ppv}
\providecommand{\latin}[1]{#1}
\makeatletter
\providecommand{\doi}
  {\begingroup\let\do\@makeother\dospecials
  \catcode`\{=1 \catcode`\}=2 \doi@aux}
\providecommand{\doi@aux}[1]{\endgroup\texttt{#1}}
\makeatother
\providecommand*\mcitethebibliography{\thebibliography}
\csname @ifundefined\endcsname{endmcitethebibliography}
  {\let\endmcitethebibliography\endthebibliography}{}

\end{document}